\documentclass[a4paper]{article}
\usepackage{INTERSPEECH}
\usepackage[latin9]{inputenc}
\usepackage{color, colortbl}
\usepackage{float}
\usepackage{multirow}
\usepackage{amsmath,bm,amssymb}
\usepackage{graphicx}
\usepackage{hyperref}
\usepackage{ragged2e}
\usepackage{microtype}
\usepackage{subfigure}
\usepackage{booktabs} 
\usepackage{tikz}
\usepackage{soul}
\usetikzlibrary{positioning}
\justifying

\newcommand\Todo[1]{ \textcolor{red}{[[ #1 ]]}}

\makeatletter

\providecommand{\tabularnewline}{\\}

\usepackage{epstopdf}
\usepackage{epsfig}
\usepackage{multicol}
\usepackage{comment}
\usepackage{url}

\title{Speaker adaptation for Wav2vec2 based dysarthric ASR}
\name{\parbox{0.9\linewidth}{\center 
Murali Karthick Baskar$^\Phi$, Tim Herzig$^\dagger$, Diana Nguyen$^\dagger$, Mireia Diez$^\Phi$, \\ Tim Polzehl$^\dagger$, Luk\'{a}\v{s} Burget $^\Phi$ and Jan ``Honza'' \v{C}ernock\'{y} $^\Phi$}  
}

\address{
$^\Phi$Brno University of Technology,  $^\dagger$Technische Universit\"at Berlin
}
\email{baskar@fit.vutbr.cz}

\begin{document}
\ninept 
\maketitle 

\begin{abstract}
Dysarthric speech recognition has posed major challenges due to lack of training data and heavy mismatch in speaker characteristics. 
Recent ASR systems have benefited from readily available pretrained models such as wav2vec2 to improve the recognition performance. 
Speaker adaptation using fMLLR and xvectors have provided major gains for dysarthric speech with very little adaptation data.
However, integration of wav2vec2 with fMLLR features or xvectors during wav2vec2 finetuning is yet to be explored.
In this work, we propose a simple adaptation network for fine-tuning wav2vec2 using fMLLR features. The adaptation network is also flexible to handle other speaker adaptive features such as xvectors.
Experimental analysis show steady improvements using our proposed approach across all impairment severity levels and attains 57.72\% WER for high severity in UASpeech dataset. We also performed experiments on German dataset to substantiate the consistency of our proposed approach across diverse domains.


\end{abstract}

\noindent\textbf{Index Terms}: Dysarthria, self-supervision, ASR, wav2vec2, fMLLR, xvectors, speaker adaptation

\section{Introduction}
 Automatic speech recognition (ASR) for dysarthric speech is hard due to two major constraints: 1) lack of sufficient data; and 2) unique speech characteristics of each speaker leading to domain mismatch.
 Data pooling~\cite{7084856} from multiple dysarthric speech datasets is one possible solution to handle the limited data issue.
 Data augmentation, GAN generated synthetic data~\cite{jin2021adversarial,vachhani2018data} and speech or volume perturbation~\cite{liu2021recent} have also been used to address the lack of sufficient data to build dysarthric ASR systems. Recent works~\cite{xiong2020source} addressed the mismatch issue~\cite{divyasree2021dad} using domain adaptation and in~\cite{wang2021unsupervised}, unsupervised domain adaptation from source to target domains is achieved using adversarial training.
 
 Finetuning a pre-trained model using supervised dysarthric speech is a widely explored technique and has shown to tackle both issues simultaneously. In addition to naive finetuning, studies have shown better gains by using auxiliary speaker information. For instance, neural network based~\cite{joy2018improving,xiong2018deep} dysarthric ASR have shown that feature space maximum likelihood linear regression (fMLLR) is a relatively superior and simple speaker normalization approach. Authors in~\cite{wang2021improved} have shown that xvectors act as complementary features to reduce speaker variability.
 
 The advent of end-to-end (E2E) ASR systems such as Transformer based ASR outperformed the hybrid systems by a significant margin. E2E has been recently applied in~\cite{wang2021improved} for dysarthric speech in combination with adaptation such as LHUC and xvectors. The authors in~\cite{yue2020autoencoder} use unsupervised bottleneck features in conjuction with fMLLR features for improving dysarthric speech recognition. Spectro-temporal deep features were used in~\cite{geng2022speaker} to remove the domain variance caused by articulatory disfluencies and perform effective speaker adaptation using multiple dysarthric datasets. In~\cite{xie2022variational}, the authors used variational autoencoder to reduce the dysarthric speech variability as an enhancement to LHUC based model adaptation. A Bayesian gated DNN architecture~\cite{liu2019use} was found to perform better acoustic  and pitch feature integration for improving robustness in dysarthric speech recognition. E2E based adaptation uses meta-learning approaches to adapt the parameters to unseen dysarthric speakers. 
 
 However, the recently proposed wav2vec2~\cite{baevski2020wav2vec}  model has shown to improve E2E systems. In this work, we attempt to finetune a wav2vec2 model in conjuction with features such as fMLLR and xvectors within a wav2vec2 framework.

 In this work, we propose a simple yet effective adaptation (finetuning) strategy  named \textit{"wav2vec2 adapter"} to incorporate wav2vec2 with speaker adaptation. Here, we attempt to finetune the wav2vec2 by feeding speaker information as auxiliary features during fine-tuning to efficiently finetune the wav2vec2 model parameters. An adapter network containing a bottleneck layer is instilled into the context encoder network of wav2vec2 model to integrate the auxiliary features and wav2vec2 outputs. The adapter module is based on the recent works~\cite{pfeiffer2020mad,pfeiffer2020adapterhub} on machine translation to fine-tune pre-trained transformer models. 
The primary contributions of this work are as follows:
\begin{itemize}
    \item We hypothesize that our adapter approach is independent of the pretraining data domain used in wav2vec2 model and empirically verify it.
    \item Wav2vec2 adapter is shown to be flexible with auxiliary features such as fMLLR and xvectors.
    \item Effect of speaker dependent and independent finetuning on wav2vec2 models are shown by comparing it with hybrid systems.
    \item Crosslingual experiments with English (EN) and German (DE) are conducted using wav2vec2 adapter framework.
    
\end{itemize}
UASpeech dataset is used to conduct our experiments and results are calculated for different severity levels. Initially, we show the importance of wav2vec2 for E2E and its comparison with hybrid systems. Experimental analysis are further conducted to substantiate the impact of our proposed wav2vec2-adapter using fMLLR and xvectors separately and jointly.
\section{Datasets used}
\textit{UASpeech} dataset is used in this work. It comprises 13 healthy control speakers and 15 dysarthric speakers. The recordings consists of isolated speech from each speaker uttering 765 words. The vocabulary includes 455 distinct words with 10 digits, 26 radio alphabets, 19 computer commands, 100 common words and 300 uncommon words. Speakers were divided into four different categories based on the severity of the condition, namely high (H), mid (M), low (L) and very low (VL). 

\textit{German dataset} is created from audio recordings of conversations between dysarthric patients and medical professionals. It contains a total of 21 dysarthric speakers, 6 females and 15 males. The corpus includes utterance level speech intelligibility ratings assessed by five naive listeners. The assessors assigned a subjective intelligibility assessment grade based on a five point scale, where "0" is assigned to clear speech and "4" being unintelligible.
The intelligibility score is used to evaluate the severity of the impairment. Based on this, the speakers were divided into four different levels of severity (high: score 3 \& 4, medium: score 2, low: score 1 and very low: score 0).\vspace{-0.2cm}
\begin{table}[ht]
  \scriptsize
  \centering
      \caption{Total amount of data in the German dataset for each dysarthric severity degree.} 
         \vspace*{-0.2cm}
  \setlength{\arrayrulewidth}{0.8pt}
  \begin{tabular}{l c c}
\toprule
Severity & Duration (MM:SS) &  Speakers \\
\midrule
Very low & 06:42 & F03, M14, M17 \\
Low  &  40:04  & F02, F05, F06, M01, M09, M12, M20  \\
Medium &  17:19 & M01, M03, M06, M10, M15 \\
High & 19:21 & F01, F04, M02, M04, M08, M16 \\
\bottomrule
  \end{tabular}
\end{table}\vspace{-0.2cm}

\begin{figure}[ht]
    \centering
    \includegraphics[scale=0.34]{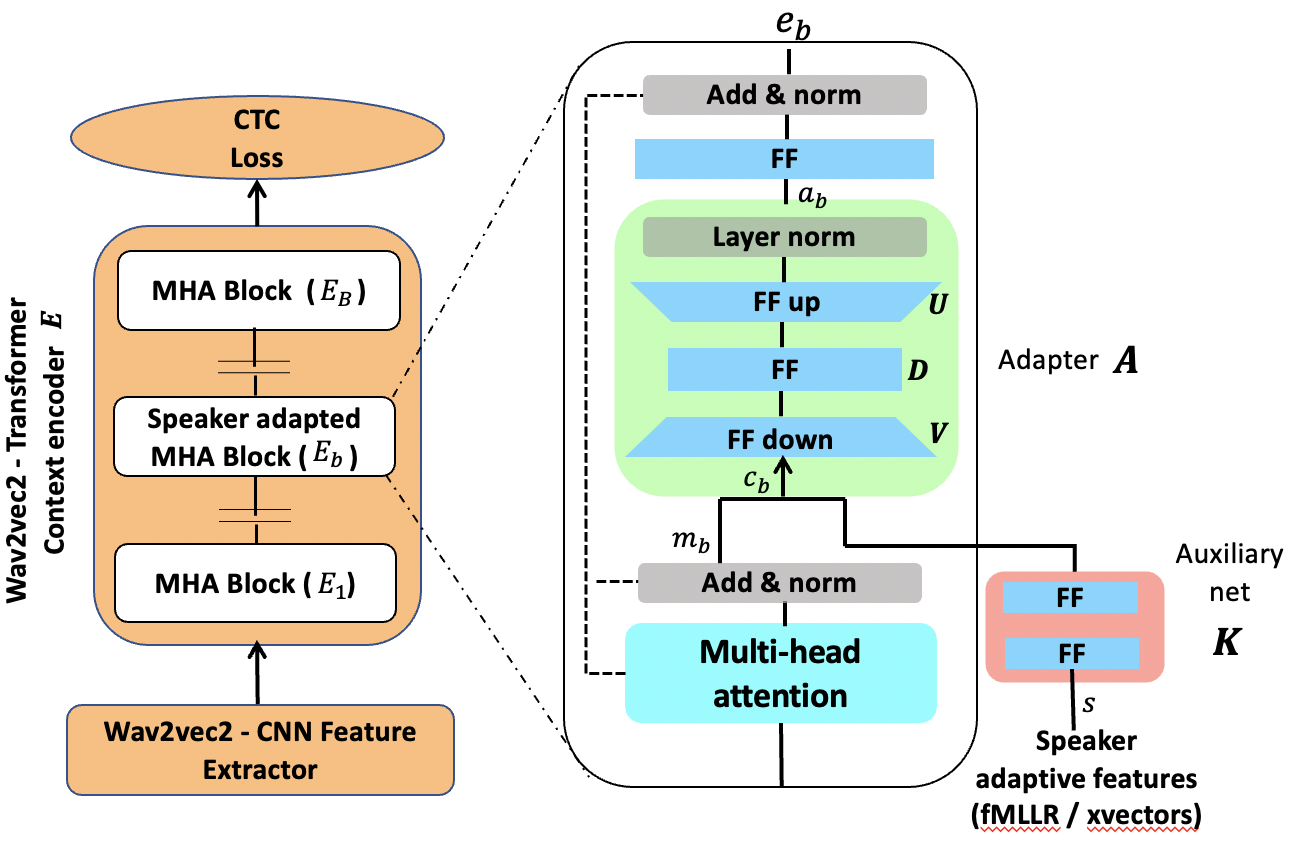}
    \vspace*{-0.2cm}
    \caption{Adaptation framework inside a pre-trained wav2vec2 model using auxiliary speaker features. The green block denotes our proposed adapter module to inherit speaker information from the auxiliary net (pink block) into wav2vec2 for performing speaker adaptation.}
    \label{fig:wife}
\end{figure}
\section{Wav2vec2 Adapter}
\subsection{Preliminaries}
The wav2vec2 are readily available pre-trained models trained with huge amounts of unsupervised data.
Let $\mathbf{X}$ be the raw input waveform. $\mathbf{X}$ is sent to the feature extractor to obtain the encoded representations.  The feature extractor contains convolutional layers and performs subsampling at a factor of 4 and directs its outputs to two parallel modules named masking component and quantizer. The masked inputs are passed into the context encoder network $\mathbf{E}$ which inherits the transformer architecture. The encoder network $\mathbf{E}$ contains $B$ multi-head attention (MHA) blocks. Each MHA block $E_{b}$ contains a MHA layer with layer norm followed by linear layer with layer norm and residual connections inbetween. 
The quantizer contains Gumbel-softmax activation which converts the unmasked inputs to quantized representations which act as labels for computation of contrastive loss training objective. In this work, we focus on adapting the parameters in the context encoder network $\mathbf{E}$ by incorporating an additional network within a block $E_{b}$ to normalize the dysarthric speaker characteristics.

\subsection{Proposed approach} 
\subsubsection{Motivation}
Wav2vec2 pretrained models have shown drastic improvements in speech recognition performance. 
Our empirical study using UASpeech on finetuning a LF-MMI~\cite{povey2016purely} and a wav2vec2 model to a particular speaker showed that LF-MMI improved in performance, while wav2vec2 did not converge. 
Besides, LF-MMI also showed further gains with fMLLR features, but speaker dependent finetuning is hard with wav2vec2. 
We hypothesize that the integration with fMLLR features can reap major benefits. We also hypothesize that incorporation of xvectors can act complementary to fMLLR features. 
\subsubsection{Model architecture}
Our proposed wav2vec2 adapter framework addresses the limitations of fine-tuning wav2vec2 in a speaker dependent manner and can be used to effectively integrate auxiliary features. The framework comprises two components: 1) an adapter; and 2) an auxiliary net. The adapter is based on the recently proposed architecture~\cite{pfeiffer2020mad} with layer normalized bottleneck layer and residual connection. Figure~\ref{fig:wife} shows the detailed block diagram view of our proposed framework. Here, we integrate the speaker adaptive features $s$ through the auxiliary net (pink block) $\mathbf{K}$ and integrate it with the adapter module $\mathbf{A}$ present in the green block. The adapter $\mathbf{A}$ is placed after the multi-head attention in a particular block $E_b$ in the wav2vec2 model. It first projects down the hidden size $h$ of the MHA block output, $m_{b}$ to $d$ dimensions using a feed-forward linear layer $\mathbf{V}$. This is followed by another $d$ dimensional linear layer $\mathbf{D}$ and projected back to $h$ using another linear layer $\mathbf{U}$. $ReLU$ activation is present after every linear layer in the architecture. The adapter $a_{b}$ output inside a particular block $E_b$, is defined as: 
\begin{align}
a_{b} &= \mathbf{U}(\mathbf{D}(\mathbf{V}(c_{b})))\\
\text{where}, c_{b} &= Append((\mathbf{K}(s)), m_{b})
\end{align}
Here, 
\begin{align}
    s &= \{f_{1}, f_{2},...,f_{T}\} \in \mathbf{F} \text{; for fMLLR features} \\
    s &= \{g_{1}, g_{2},..,g_{T}\}\in\mathbf{G} \text{; for xvectors} 
\end{align}
$f_{t}$ denotes a fMLLR feature at time $t$ from a sequence $\mathbf{F}$ and xvectors $G$ is computed for each utterance $u$ to get $g_{u}$ and is repeated for all time steps $T$ to get $g_{1:T}=g_{u}$.
Unlike the adapter net in machine translation~\cite{pfeiffer2020mad} which uses an adapter after every block, we only use a single adapter module after a particular block in wav2vec2. This setup helps to avoid drastic increase in parameters. We also use an auxiliary net and integrate it to the input of the adapter for instilling speaker adaptation during finetuning.
Our proposed wav2vec2 adapter is flexible to integrate any type of auxiliary information to perform speaker normalization.
\subsubsection{Auxiliary Features}
In this work, we explore the wav2vec2 adapter module using supervised speaker adaptive features "fMLLR" and unsupervised speaker vectors "xvectors". These are described as follows:

\textit{fMLLR}: The fMLLR features are obtained from the MLLR transforms trained with supervised adaptation data. Conventional fMLLR transformation is applied over MFCC or filterbank features. In our work, we extract input features for computing fMLLR transforms from the output (1024 dims) of the final block $E_{B}$ in figure~\ref{fig:wife} by forward propagating the raw waveforms as wav2vec2 input. We arrived to the usage of wav2vec2 based features for fMLLR computation and its integration at the $B\textsuperscript{th}$ MHA block empirically. 
The collected 1024 dim. wav2vec2 features are transformed using LDA followed by GMM-HMM based training.
Kaldi recipe\footnote{https://github.com/kaldi-asr/kaldi/blob/master/egs/wsj/s5/run.sh\#L211}
is used for LDA+MLLT+SAT training and speaker dependent fMLLR transforms computation to get 40 dimensional fMLLR features $\mathbf{F}$.

\textit{XVectors}: Xvectors $\mathbf{G}$ are extracted from MFCC features using a separately trained xvector extractor. Since the xvectors are not directly extracted from wav2vec2 features, we integrate the xvectors with the output of the intermediate block i.e. $E_{b}$ in figure~\ref{fig:wife}, here $b=2$ is decided empirically (see section \ref{sec:MHAblockxvector}). To substantiate our results, we follow a similar strategy to the one presented in~\cite{karafiat2016multilingual},  where the authors integrated ivectors for neural network adaptation in~\cite{karafiat2016multilingual}.

Finetuning with wav2vec2 adapter involves two stages: 1) the model parameters within the green and pink block are initially  updated by fixing the rest of parameters. This first update helps the wav2vec2 encoder establish connections with the auxiliary information. 2) The whole model is then finetuned till convergence. Training configuration is explained in section~\ref{sec:es}.

\section{Experimental setup}~\label{sec:es}

\textit{Architectures and Initialization}:  Kaldi~\cite{povey2011kaldi} is used to train our baseline DNN-HMM, LF-MMI hybrid systems and the fMLLR feature extraction. Transformer and Wav2vec2 systems are built using ESPnet toolkit~\cite{watanabe2018espnet}. DNN-HMM contains 7 hidden layers each with 2048 dimensions. LF-MMI is built with 13 TDNN layers trained using sequence-level objective with data augmentations and ivectors. Our implementations will be available\footnote{https://github.com/creatorscan/Dysarthric-ASR} with updates. 

Transformer model contains 24 encoder blocks and 6 decoder layers each with 1024 dimensions. The transformer is initially pretrained with 960 hours of Librispeech data as we found random initialization did not lead to convergence with dysarthric data due to increased model complexity. 
Wav2vec2 follows similar architecture as the transformer and in this work we use two types of models: 1) wav2vec2 pretrained with 60k hours of Libri-light data and finetuned with 960 hours of Librispeech data~\footnote{https://huggingface.co/facebook/wav2vec2-large-960h-lv60}. Experiments on UASpeech are conducted using this wav2vec2 model. 2) XLSR wav2vec2 pretrained with 53 multilingual languages ($\approx$ 56k hours  from the MLS, CommonVoice and Babel corpora) as defined in~\cite{conneau2020unsupervised}. 2k hours of German data from CommonVoice is later used to finetune the XLSR~\footnote{https://huggingface.co/facebook/wav2vec2-large-xlsr-53-german}. 

\textit{Training configuration}:
The wav2vec2 model is finetuned by optimizing using Adam with weight decay fix \cite{loshchilov2017decoupled} with the parameters: $\beta_1$ = 0.9, $\beta_1$ = 0.999, $\epsilon$ = 1e-8 and $L_2$ weight decay of 0. We set an initial learning rate of 1e-4. The learning rate is warmed up over the first 500 steps then linearly decayed. We choose a total number of training epochs of 30. The global norm of the gradient is clipped at 1.0. 

\textit{xvector extraction}:
512 dimensional xvectors are extracted from a multilingual (34 languages) extractor trained with 1.65M recordings from 78k speakers using CNN-TDNN architecture. The detailed description of xvector extraction are in~\cite{karafiat2021analysis}.

\section{Results and discussion}
We intially perform an experimental study on different model architectures such as DNN-HMM, LF-MMI and Transformer with fMLLR and model adaptation. The model adaptation (MA) is done by fine-tuning all the model parameters of DNN-HMM or LF-MMI in two stages: During stage 1, the model is trained with all supervised dysarthric data independent of speakers. The trained model is adapted to each speaker at stage 2.
Incase of Transformer MA, the model parameters are initialized with 960hrs of Librispeech, which is followed by speaker independent re-training with dysarthric data. We found randomly initialized Transformer did not result in convergence. Speaker dependent training also degraded the Transformer model performance which is in contrary to Hybrid models. 

\subsection{Effect of fMLLR features}
\begin{figure}[ht]
    \centering
    \includegraphics[trim={0cm 0.2cm 0.2cm 0cm},clip,scale=0.32]{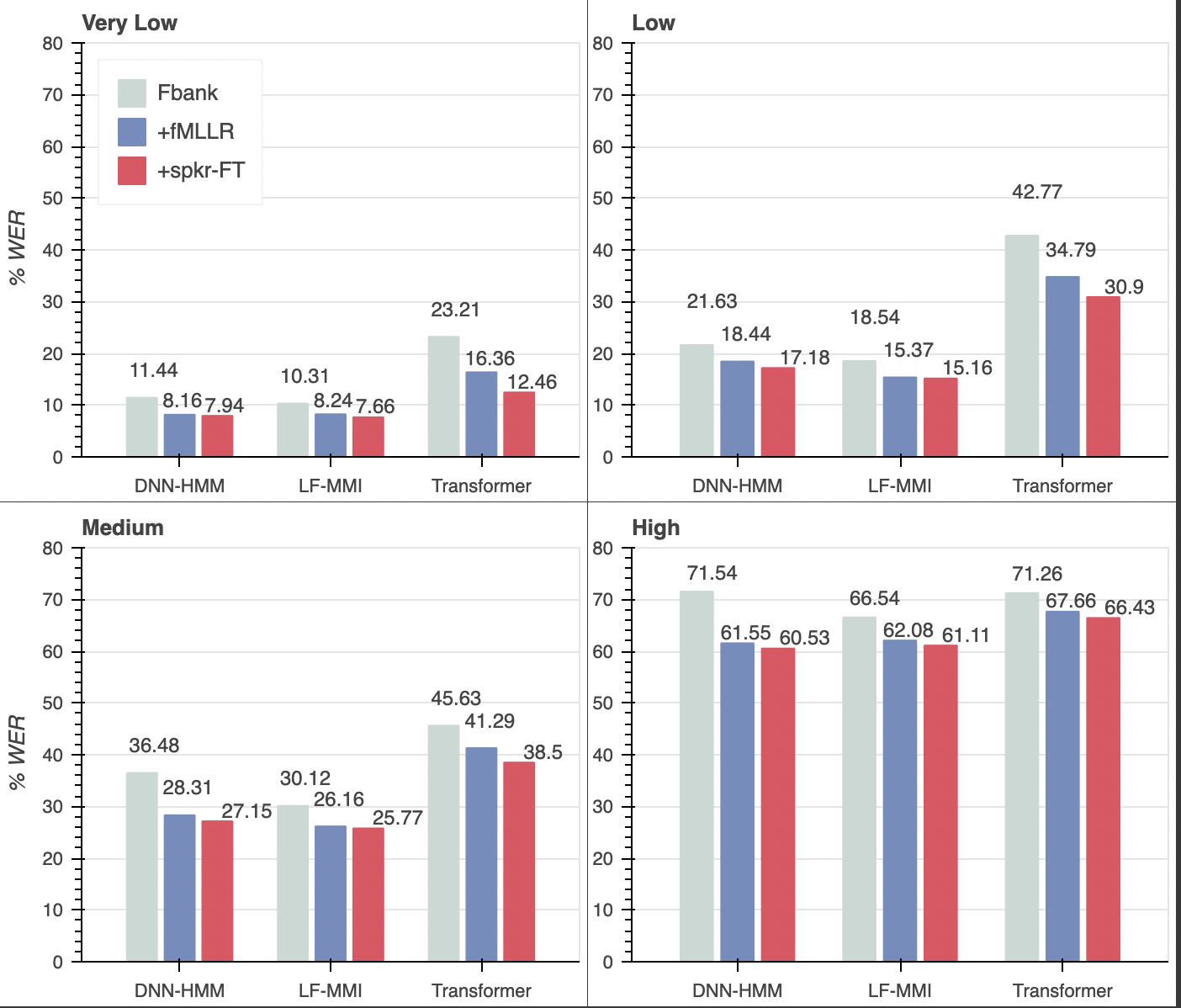}
    \vspace*{-0.1cm}
    \caption{Performance comparison of fMLLR features with speaker level finetuning (spkr-FT) across different model architectures such as DNN-HMM, LF-MMI and Transformers}
    \label{fig:analysis}
\end{figure}
DNN-HMM, LF-MMI and Transformers are initially trained with fMLLR features of all speakers  and later finetuned to each speaker.
Figure~\ref{fig:analysis} compares the models finetuned with and without fMLLR features.
Among hybrid models, LF-MMI outperforms DNN-HMM in all cases. Transformer based ASR did not show any improvement when compared to either of the hybrid systems. Nevertheless, finetuning with fMLLR features has consistently outperformed the baseline filterbank based systems. 

To improve the capability of Transformer, wav2vec2 model is used on behalf of transformer. 
Results are in table~\ref{tab:fmllr_xvec}. 
Unlike the Transformer system, Wav2vec2 finetuned with dysarthric data shows comparable performance to LF-MMI. Finetuning wav2vec2 with our adapter module using fMLLR features significantly outperforms our best LF-MMI system.
\begin{table}[ht]
\caption{WER\% results for our wav2vec2 adapter module, finetuned (FT) with fMLLR features, with xvectors and with their combination}\label{tab:fmllr_xvec}\vspace*{-0.2cm}
\centering
\scriptsize
\begin{tabular}{clcccc}
\toprule 
Model & Adaptation & VL & L & M & H\tabularnewline
\hline 
\midrule
\multirow{1}{*}{LF-MMI} & fMLLR+Spkr-FT & 7.66  & 15.16  & 25.77 & 61.11 \tabularnewline
\midrule
  \multirow{4}{*}{Wav2vec2} & FT & 8.54 & 16.67  & 33.70 & 64.78 \tabularnewline
& fMLLR+FT & 7.42 & 12.07 & 23.11 & 58.48\tabularnewline
& xvectors+FT & 7.11 & 13.42 & 23.52 & 59.52\tabularnewline
& xvectors+fMLLR+FT & 6.83 & 12.82 & 22.46 & 57.72 \tabularnewline
\bottomrule
\end{tabular}
\end{table}\vspace{-0.2cm}

\subsection{Optimal MHA block for xvector Integration}
\label{sec:MHAblockxvector}
\begin{figure}[ht]
    \centering
    \includegraphics[scale=0.4]{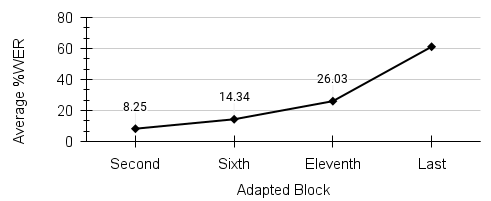}
    \vspace*{-0.2cm}
    \caption{Performance comparison of integrating xvectors at different blocks inside wav2vec2}
    \label{fig:xvectors}
\end{figure}

Figure~\ref{fig:xvectors} shows the analysis to find the optimal block inside wav2vec2 encoder to integrate xvectors. 
Figure~\ref{fig:xvectors} shows a clear trend that feeding xvectors at higher level blocks resulted in performance degradation in terms of average \%WER across all speakers. 
Therefore, based on empirical evidence, we select the second block $E_2$ to be used as an adapter module to integrate xvectors (unlike fMLLR features which are fed at the last block). 

Table~\ref{tab:fmllr_xvec} shows the importance of adapting the wav2vec2 using fMLLR and xvectors. fMLLR features shows clear wins compared to xvectors for all severities except the VL severity. Adapting the wav2vec2 with two adapters, one placed at first block fed with xvectors and the second adapter in the last block fed with fMLLR features show complementarity with each other by attaining consistent gains.

\begin{table}[ht]
\caption{Crosslingual experiments on English (EN) and German (DE) with model adaptation}
\vspace*{-0.2cm}\label{tab:xl}
\centering
\scriptsize
\begin{tabular}{ccccc}
\toprule 
Type & Training & Finetuning & Evaluation & Avg. \%WER \tabularnewline
\hline 
\midrule
\multirow{2}{*}{Baseline} & EN & - & EN & 24.21   \tabularnewline 
& DE & - & DE & 28.43  \tabularnewline 
\midrule
\multirow{4}{*}{Crosslingual} &\multirow{4}{*}{EN+DE} & - & EN & 24.25  \tabularnewline
& & - & DE  & 27.32 \tabularnewline
 & & EN & EN & 23.76 \tabularnewline
& & DE & DE & 25.93   \tabularnewline
\bottomrule
\end{tabular}
\end{table}\vspace{-0.2cm}
\subsection{Cross-lingual experiments on English and German}
A crosslingual (XLSR) model is initially built by combining EN from UASpeech and DE from our German corpus. The output layer contains 75 dimensions by combining 40 phonemes from English and 35 phonemes from German along with their corresponding language tag. The output layer is connected to the pre-trained XLSR model followed by fine-tuning with both EN and DE together and the resulting model is evaluated for EN and DE. The results in table~\ref{tab:xl} show a 4.2\% relative improvement for DE over the baseline system but no improvement for EN. This is because DE has very little training examples compared to EN, which makes the crosslingual training beneficial for DE. 
Further gains are obtained by finetuning to each language independently and the results are in table~\ref{tab:xl}.

\begin{table}[ht]
\caption{Performance of wav2vec2 adapter, CUHK system on 3 impaired speakers in UASpeech}
\vspace*{-0.2cm}
\label{tab:cuhk}
\centering
\scriptsize
\begin{tabular}{clcc}
\toprule 
ID & F04 (Low) & M12 (High) & M14 (VeryLow)\tabularnewline
\hline 
\midrule
Dysarthria & Athetoid CP & Mixed  & Spastic CP \tabularnewline
CUHK~\cite{hu2019cuhk} & 24.0 & 54.0 & 15.6 \tabularnewline
Ours & 22.5 & 49.9 & 12.9   \tabularnewline
\bottomrule
\end{tabular}
\end{table}
\subsection{Comparison across Literature}
Among several works~\cite{wang2020improving,wang2021improved,hu2019cuhk,liu2021recent} on UASPeech, authors in
CUHK-2019~\cite{hu2019cuhk}, showed that lip features from video helps to improve the hybrid DNN system.
In CUHK-2021,~\cite{liu2021recent} performed neural architecture search (NAS) and combined it with data augmentation (DA), video embedding (AV) and LHUC based adaptation. In 2022,~\cite{geng2022speaker} showed that hybrid DNN trained with spectro-temporal features and xvectors outperformed their Conformer system. Spectro-temporal features capture longer context for better normalization of speaker variability and hence improves over CUHK system.
In table~\ref{tab:cuhk} we show how our proposed system behaves and improves over CUHK system~\cite{hu2019cuhk,wang2021improved} across different types of dysarthric speakers. 
Table~\ref{tab:pub} shows the average \%WER across all severity levels between existing best systems.   Our proposed wav2vec2 adapter and its multilingual extension attains 3.35\% and 5.15\% relative improvement over SBE~\cite{geng2022speaker} respectively. 
\begin{table}[h]
\caption{A comparison between published works on UASpeech and our proposed work}
\vspace*{-0.2cm}
\label{tab:pub}
\centering
\scriptsize
\begin{tabular}{cc}
\toprule 
Systems & Avg. \%WER \tabularnewline
\hline 
\midrule
CUHK-2021 NAS DNN+ DA + LHUC + AV~\cite{wang2021improved} & 25.21\tabularnewline
DA + SBE + LHUC~\cite{geng2022speaker}  & 25.05   \tabularnewline 
Wav2vec2 +fMLLR+xvectors (Ours) & 24.21  \tabularnewline
Wav2vec2 (XLSR) +fMLLR+xvectors+Mult. (Ours)  & 23.76  \tabularnewline
\bottomrule
\end{tabular}
\end{table}\vspace{-0.2cm}
\section{Conclusion}
In this work, we investigate the importance of speaker dependent auxiliary features such as fMLLR and xvectors for adapting wav2vec2 models for improving dysarthric speech recognition. We show that in contrast to hybrid systems, wav2vec2 did not improve by adapting model parameters  based on each speaker. 
We proposed a wav2vec2 adapter module that inherits speaker features as auxiliary information to perform effective speaker normalization during finetuning. 
We showed that, using the adapter module, fMLLR and xvectors are complementary to each other, and proved the effectiveness of the approach outperforming existing SoTA on UASpeech dysartric speech ASR.
In our cross-lingual experiments, we also showed that combining English and German data for training, can further improve performance of our systems, proving useful in scenarios where little training examples exist for a particular language.

\section{Acknowledgements}
The work was supported by European Union's Horizon 2020 project HumaneAI-Net, No. 761758, Czech National Science Foundation (GACR) project NEUREM3 No. 19-26934X, and Czech Ministry of Education, Youth and Sports project no. LTAIN19087 "Multi-linguality in speech technologies". Computing on IT4I supercomputer was supported by the Czech Ministry of Education, Youth and Sports from the Large Infrastructures for Research, Experimental Development and Innovations project "e-Infrastructure CZ-LM2018140".
\bibliographystyle{ieeetr}
\bibliography{ref_new}

\end{document}